\documentclass[11pt]{article}
\usepackage{fullpage}
\usepackage{amsmath}
\usepackage{amssymb}
\usepackage[dvips]{epsfig}

 \advance \textwidth by -2in
 \advance \textheight by -3in
\def\##1{\underline #1}
\def\=#1{\underline{\underline #1}}

\begin{document}

\begin{center}
{\large {\bf On Direct and Indirect Scattering Approaches for Homogenization
of Particulate Composites}}
\vskip 0.4cm

\noindent  {\bf Akhlesh Lakhtakia\/}\footnote{Tel: +1 814 863 4319; Fax: +1 814 863 7967; e-mail: AXL4@psu.edu\\} 
\vskip 0.2cm
\noindent {\em CATMAS~--~Computational \& Theoretical Materials Sciences Group\\
Department of Engineering Science \& Mechanics\\
Pennsylvania State University, University Park, PA 16802--1401, USA}

\end{center}

\vskip 1.0cm

\noindent {\bf Abstract:} Theoretical formalisms 
for the homogenization of 
particulate composites are identified as following
either  the direct scattering approach (DSA) or the
indirect scattering approach (ISA). Both approaches
can take inclusion size--dependence and distribution
statistics into account.
However, the DSA is generally limited to mediums
with direction--independent constitutive properties and
inclusions with simple shapes, but the ISA is not hobbled
thus.

\vskip 0.2cm


\newpage

\section{Introduction}

Random dispersions of identical, electrically small inclusions in
a homogeneous host medium can be equivalently
considered as {\it effectively\/} homogeneous, after using
various homogenization formalisms \cite{LOCM}. Two broad 
approaches can be discerned from the homogenization literature 
for particulate composites,
spanning almost two centuries.
Any inclusion is considered as a scatterer,
in either  approach. In the {\it direct scattering
approach\/}, DSA for short, the relevant scattering problem is treated
as a boundary value problem. This is exemplified by the 
multiple scattering theory \cite{Lax, WT}\footnote{The homogenization literature
is vast, and only  representative
works are cited here. Unless mentioned otherwise, attention is focused
here on 3D composites.} and the structural unit
model \cite{PS}. In the {\it indirect scattering approach\/} (ISA),
on the other hand, a scatterer is treated as a source of an {\em excess\/}
electromagnetic field in some embedding medium; and 
the field induced in an electrically small, exclusion region
of the same shape (and, possibly, size) as the inclusion has to be estimated.
Derivations of the Maxwell Garnett
formula and its size--dependent variants from integral
equations \cite{Lopt,SS99} furnish suitable examples.

Both approaches may be equivalent for the homogenization of simple composites.
For example, Maxwell Garnett \cite{MG}
applied a DSA
to obtain his eponymous formula for
isotropic, dielectric--in--dielectric, particulate
composites with spherical inclusions~---~towards which
formula an ISA was provided later
by Fax\'en \cite{Fax}. Likewise,  for the
same type of composites,
Bruggeman \cite{Br} adopted the DSA to get the Bruggeman formula,
which also emerges naturally from an ISA involving
 the bilocal approximation
in the context of the strong property fluctuation theory (SPFT) \cite{TK81}.
When the DSA is adopted, 
the original Maxwell Garnett and the
Bruggeman
formulas can be extended to cover  chiral--in--chiral, particulate
composites with spheroidal inclusions (as will become clear in the next section).
But in a framework of the ISA, extensions of the same formulas have been made
for much more
complex~---~i.e., anisotropic and 
bianisotropic~---~composites with inclusions of aciculate, discotic, ellipsoidal and
other convex shapes \cite{LMW97,WLM97}.

The foregoing delineation of the two scattering approaches to homogenization
has never been published earlier, to the author's knowledge. Consequently, no 
comparison of their different capabilities has every been reported. That situation
is remedied in this brief review, mostly with reference to particulate composites with
complex constitutive properties.

\section{Direct Scattering Approach}

Adoption of the DSA requires that a boundary value problem be solved
for scattering by an inclusion in some embedding medium. The embedding medium is,
e.g., the host medium in the Maxwell Garnett formalism, and
the homogenized composite in the Bruggeman formalism.

The incident field
and the scattered field must be expanded in terms of appropriate
eigensolutions of the
frequency--domain Maxwell postulates applied to the
embedding medium. In practical terms, the field
induced inside the inclusion must also be expanded in terms of
appropriate eigensolutions of the Maxwell postulates applied
to the inclusion medium, which is best seen in the T--matrix method
and its simplifications for inclusions of special geometries \cite{LBel,MVV}.
Boundary conditions across the 
surface of the scatterer are then enforced, and the scattered field coefficients
are related to the incident field coefficients. Then, either the scattered field
is made to satisfy some condition \cite{Boh,Doyle}, or
the total field everywhere is 
averaged in some fashion \cite{Lax,WT,MVV}.

Clearly, if appropriate eigensolutions are not available, the DSA turns out
be barren. At this time, the following remarks on the utility
of the DSA are in order:
\begin{itemize}
\item Both the inclusion and the embedding mediums must have
direction--independent constitutive properties \cite{LBel}. Consistently
with the Post constraint \cite{Post}, this means that the most complex 3D
homogenization
problem thus solvable can have isotropic chiral inclusions embedded in
an isotropic chiral medium \cite{Appa, Shan}. 

\item For 2D problems wherein the inclusions are parallel,
infinitely long cylinders with convex cross--sections, 
both the inclusion and the embedding mediums are 
allowed to have
uniaxial constitutive dyadics, provided the crystallographic axes
of both mediums are parallel to the inclusion axis. Although no
results appear to have been reported, the situation described should be mathematically
tractable \cite{Yin}. Thus uniaxial bianisotropic materials may fall within the purview
of the DSA, to a limited extent.

\item The DSA is most tractable when the inclusions are either spheres (3D problems)
or infinitely long cylinders with circular cross--sections (2D problems). These
restrictions can be relaxed somewhat because analytic continuation
is invoked
in the T-matrix methodÊ\cite{LBel, IBDM, ILD}, 
so that spheroids and cylinders with elliptical cross--sections
may also be entertained to a limited degree. The T--matrix
method for scatterers of other shapes is not generally practical.
Alternately, by
virtue of the helicity properties of Beltrami fields \cite{Trk},
the use of spheroidal eigenfunctions (3D problems) 
\cite{AY} and elliptical eigenfunctions (2D problems) \cite{BW}
may be helpful if both the inclusion and
the embedding mediums have direction--independent properties.

\end{itemize}

A major advantage of DSA is that it allows the use of inclusion distribution
statistics, such as the pair--correlation function \cite{Lax,WT,MVV,Tw}. Furthermore,
as exemplified by Doyle \cite{Doyle} and Ma {\it et al.\/} \cite{MVV}, the DSA can take
the actual size of each inclusion into account, in addition to the volumetric
fraction of the inclusion medium. But, by no means are both advantages
unique to the adoption of the DSA.

\section{Indirect Scattering Approach}

The ISA is less concerned with the scattering response of a solitary
inclusion as with the estimation of the source--region
field, i.e., the field induced inside
a so--called {\em exclusion region\/} which is of the same shape (and size) as the inclusion.
This estimate can be obtained, once an embedding medium is decided upon.
The embedding medium can be conveniently chosen, e.g., in the
SPFT \cite{ZL99}.

When the dyadic Green function for the embedding medium is
explicitly known, the source--region field can be estimated for
any convex--shaped exclusion region, and the exclusion region
need not be of infinitesimal dimensions \cite{WLsr1,WLsr2}. Otherwise,
the source--region field can be estimated for an arbitrary
bianisotropic medium, provided the exclusion region is ellipsoidal
and is treated as infinitesimally small
\cite{MW}. Thus, the ISA enables the homogenization of
composites that have far more complex properties~---~in terms of constitutive
properties as well as inclusion shapes~---~than the DSA does.

Whereas the Maxwell Garnett and the Bruggeman formalisms
take only the volumetric fraction of the inclusion medium
into account, higher--order statistical measures of the
distribution of inclusions can be incorporated in the SPFT \cite{TK81,Z94,ML95}.
In this attribute, the SPFT is at par with the multiple
scattering theory \cite{Lax,WT,MVV,Tw}.

\section{Concluding Remarks}

To conclude, two scattering approaches for the homogenization of
particulate composites have been identified here.
Both approaches have been useful in the past
and will continue to be so. Although inclusion
size--dependence as well as inclusion distribution
statistics can be addressed in both approaches,
the
scope of the ISA is much more vast than of the DSA.
To a large degree, the DSA is confined to mediums
with direction--independent constitutive properties and
inclusions with simple shapes; but the ISA is not hobbled
in that manner. Let us note, in this connection, that the
scope of either approach can be enhanced somewhat
by using such mathematical devices as the affine transformation \cite{Berth,SSBM}
and the addition of gyrotropic--like magnetoelectric components 
to the constitutive
dyadics \cite{gyr}.  The variety of nonlinear composites that can be treated
with the DSA \cite{ZBHS,SAM} is considerably limited than those treatable with the ISA
\cite{LWn1,LWn2},
because nonlinearity normally appears hand--in--hand with anisotropy
but rarely with isotropy \cite{Kob,Boyd}. Finally,
nonlocality in homogenized composites can also be tackled 
using the ISA \cite{ZL99,ML95}.

\end{document}